\documentclass[twocolumn]{aastex6}

\usepackage{graphicx}
\usepackage[space]{grffile}
\usepackage{latexsym}
\usepackage{textcomp}
\usepackage{gensymb}
\usepackage{multirow,booktabs}
\usepackage{amsfonts,amsmath,amssymb}
\usepackage{natbib}
\usepackage{url}
\usepackage{hyperref}

\newif\iflatexml\latexmlfalse
\DeclareGraphicsExtensions{.pdf,.PDF,.png,.PNG,.jpg,.JPG,.jpeg,.JPEG}

\usepackage[utf8]{inputenc}
\usepackage[english]{babel}
\usepackage{CJK}

\def\gtrapprox  {\;\lower 0.5ex\hbox{$\buildrel >\over \sim\ $}}
\def\lessapprox {\;\lower 0.5ex\hbox{$\buildrel < \over \sim\ $}}
\def\HI{\text{H\,\textsc{I}}~}   
\def\SiIII{\text{Si\,\textsc{iii}}}
\def\SiII{\text{Si\,\textsc{ii}}}
\def\SiIV{\text{Si\,\textsc{iv}}}
\def\OI{\text{O\,\textsc{i}}}

\def\CII{\text{C\,\textsc{ii}}}
\def\CIV{\text{C\,\textsc{iv}}}
\def\NV{\text{N\,\textsc{v}}}
\def\FeII{\text{Fe\,\textsc{ii}}}
\def\AlII{\text{Al\,\textsc{ii}}}
\def\msun{\mbox{$\rm M_\odot$}}
\def\lya{Ly$\alpha$~}
\def\nhi{\mbox{$N_{\rm HI}$}}
\def\h100{\mbox{$h^{\rm -1}$}}
\def\kms{\mbox{$\rm ~km~s^{-1}$}}
\def\cm-2{\mbox{$\rm ~cm^{-2}$}}
\def\mA{\mbox{m\AA}}
\def\Wprime{W$^{\prime}$~}

\def\rvir{$R_{\rm vir}$}

\def\sigmav{$\sigma_{\rm v}$}

\DeclareGraphicsExtensions{.pdf, .ODF, .eps}
\shorttitle{\lya Absorbers and the Coma Cluster}
\shortauthors{Yoon \& Putman}
\slugcomment{11/2016}

\begin{document}
\begin{CJK*}{UTF8}{}

\title{\lya Absorbers and the Coma Cluster}

\author{Joo Heon Yoon (\CJKfamily{mj}윤주헌) \& M.E. Putman$^1$}
\affil{$^1$ Department of Astronomy, Columbia University, New York, NY 10027, USA}

\begin{abstract}
The spatial and kinematic distribution of warm gas in and around the Coma Cluster is presented through observations of \lya absorbers using background QSOs. Updates to the \lya absorber distribution found in \cite{Yoon2012a} for the Virgo Cluster are also presented.  At 0.2-2.0\rvir\ of Coma we identify 14 \lya absorbers (\nhi$ = 10^{12.8-15.9}\cm-2$) towards 5 sightlines and no \lya absorbers along 3 sightlines within 3\sigmav$_{\rm coma}$.
For both Coma and Virgo, most \lya absorbers are found outside the virial radius or beyond 1\sigmav\ consistent with them largely representing the infalling intergalactic medium. 
The few exceptions in the central regions can be associated with galaxies. 
The \lya absorbers avoid the hot ICM, consistent with the infalling gas being shock-heated within the cluster.  The massive dark matter halos of clusters do not show the increasing column density with decreasing impact parameter relationship found for the smaller mass galaxy halos.   In addition, while the covering fraction within \rvir\ is lower for clusters than galaxies, beyond \rvir\ the covering fraction is somewhat higher for clusters.  The velocity dispersion of the absorbers compared to the galaxies is higher for Coma, consistent with the absorbers tracing additional turbulent gas motions in the cluster outskirts.
The results are overall consistent with cosmological simulations, with the covering fraction being high in the observations standing out as the primary discrepancy.

\end{abstract}

\section{Introduction}

Baryons accrete onto dark matter halos throughout the universe. Galaxy clusters represent the most massive case and therefore the gaseous extremes of large velocities, strong shocks, and high temperatures. 
Once the baryons are within the virial radius of a cluster, they are hot enough to be mapped out in detail due to their high X-ray emissivities.  This is very different from typical galaxies where the hot baryons are nearly invisible.  
Though there are many differences between high (cluster) and low (galaxy) mass halos, numerical simulations predict the accretion from the intergalactic medium (IGM) is filamentary and multi-phase in both cases.  

A significant percentage of the accreting gas from the IGM is expected to be warm ($10^{4-5}$ K; \cite{kravtsov12,joung12,Avestruz2016a} \citet*[][hereafter EBP15]{Emerick2015a}). Although clusters have been extensively studied, there is very little observational data on the presence of warm diffuse gas in and around them.  This is unlike individual galaxies, which have been the subject of extensive surveys of their circumgalactic environment \citep[e.g.,][]{Tumlinson2013a, Chen2001a}.
The warm gas in and around clusters traces the dynamic and thermal state of the infalling gas, and therefore tests our theoretical model of cluster formation \citep[e.g.,][]{moore96,loken02,voit05}.  Its relation to other phases also constrains how gas cooling and heating are treated in cosmological simulations \citep[e.g.,][]{kravtsov05}.  Knowledge of the warm gas distribution strengthens our understanding of cluster physics, and has the potential to affect the cosmological constraints obtained when using cluster surveys to measure dark energy \citep[e.g.,][]{nagai14,lau09,younger06}.

In this paper, we map the distribution and kinematics of the warm gas in and around the Coma Cluster using the Cosmic Origins Spectrograph (COS) on HST to observe Ly$\alpha$ absorbers towards multiple sightlines. 
Coma is a massive ($\sim1.4\times10^{15}$ M$_{\odot}$), regular cluster at $\sim100$ Mpc \citep{Lokas2003a,Colless1996a}.   It is the closest rich cluster \citep[Abell Class 2; ][]{Abell1989a} and has therefore been the target of many multiwavelength studies, including an HST/ACS Treasury Survey \citep{hammer10}, large Chandra surveys (PI: Sanders), deep GALEX UV observations \citep{smith10}, and extensive HI and H$\alpha$ surveys \citep[e.g.,][]{Bravo-Alfaro2000a,yagi10}. These studies have found Coma to be a rich cluster with thousands of galaxies and clear infalling substructures \citep{Colless1996a}. The Coma Cluster is considered a template for our understanding of structure formation in the Universe \citep{zhuravleva13,peebles70,gott71}.

\begin{deluxetable}{cccc}
\tabletypesize{\scriptsize}
\tablecaption{\label{property.tab} The properties of the Coma and Virgo Clusters}
\tablewidth{0pt}
\tablehead{
  \colhead{} &
  \colhead{Coma} &
  \colhead{Virgo} & 
  \colhead{ref.} }
\startdata
$M_{\rm vir}$ (\msun) & $1.4\times10^{15}$ & $2.19\times10^{14}$ & a,b \\
$R_{\rm vir}$ (Mpc) & 2.9 & 1.57 & a,b \\
D (Mpc) & 105 & 16.5 & c,d \\
$v$ (\kms) & 6930 & 1138 & e,d \\
$\sigma_{\rm v}$ (\kms) & 1008 & 544 & e,d 
\enddata
\tablecomments{a: \citet{Lokas2003a} b: \citet{mamon04} c: \citet{Ferrarese2000a} d: \citet{Mei2007a} e: \citet{Struble1999a}}
\end{deluxetable}

This study of Coma permits us to examine how the warm gas properties of a cluster change with mass.  In particular, in \cite{Yoon2012a} (hereafter Y12), we presented a map of absorbers in and around the less massive, irregular Virgo Cluster, and here we present a map updated with new sightlines.  In Y12 we found the warm gas seemed to avoid the hot ICM, and traced infalling subgroups of galaxies and warm gas.  
Coma is much more settled than Virgo, but it does have evidence for infalling substructure from the galaxy dynamics and morphologies and a radio relic that may be an infall shock \citep{Brown2011a}. It is $\sim6-7$ times the mass of Virgo, and may have significantly larger flows of warm gas into the cluster.  
On the other hand, there have been suggestions that baryons accreting onto massive clusters are pre-processed through groups, and this may decrease the amount of warm gas \citep{wetzel12}.  
In the central regions, we may expect there to be little warm gas, as found for Virgo, because the ICM is even hotter and heat conduction will be more important.     

This paper begins by describing the COS observations for 8 QSOs that lie behind the Coma Cluster within 2 times the virial radius (\S2).   This section also presents the new sightlines that pass within $\sim2.5$\rvir~of the Virgo Cluster.  The results of the observations are presented in \S3, including the spatial and velocity distribution of the absorbers, the relationship of the absorbers to cluster galaxies, and the covering fraction of the gas.  In \S4 the results are discussed in the context of the mass and structure of the clusters, the physical processes in galaxies, the kinematics of the absorbers in the context of tracing turbulence, and recent simulations of gas flows relative to a Coma-like and Virgo-like cluster (EBP15).  The main physical properties adopted for Coma and Virgo in this paper are shown in Table~\ref{property.tab}.  In particular, the distance to the Coma Cluster is assumed to be 105 Mpc throughout \citep{Ferrarese2000a}.

\medskip
\section{Observations and Data Reduction}
\label{obs.sec}

The observations were conducted for 9 QSO sightlines within 2\rvir\footnote{We are adopting the virial radius of \cite{Lokas2003a} which is defined to be 2.9 Mpc where the density is $\sim100$ times the current critical density.} of the Coma Cluster using HST/COS in medium resolution mode with the G130M grating in Cycle 21 (PID 13382; PI Putman).
All QSO sightlines were observed using the 1291 \AA\ and 1309 \AA\ central wavelengths and we employed 2 FP-POS positions to dither the spectra in order to reduce the impact of fixed-pattern and grid-wire flat field effects.  The spectrum of one sightline, J1252+2913, was wiped out by a Lyman limit system (LLS) at a higher redshift than the Coma Cluster, and therefore only 8 sightlines were used in this study.

\begin{figure*}[ht!]
\begin{center}
  {\includegraphics[width=1.7\columnwidth]{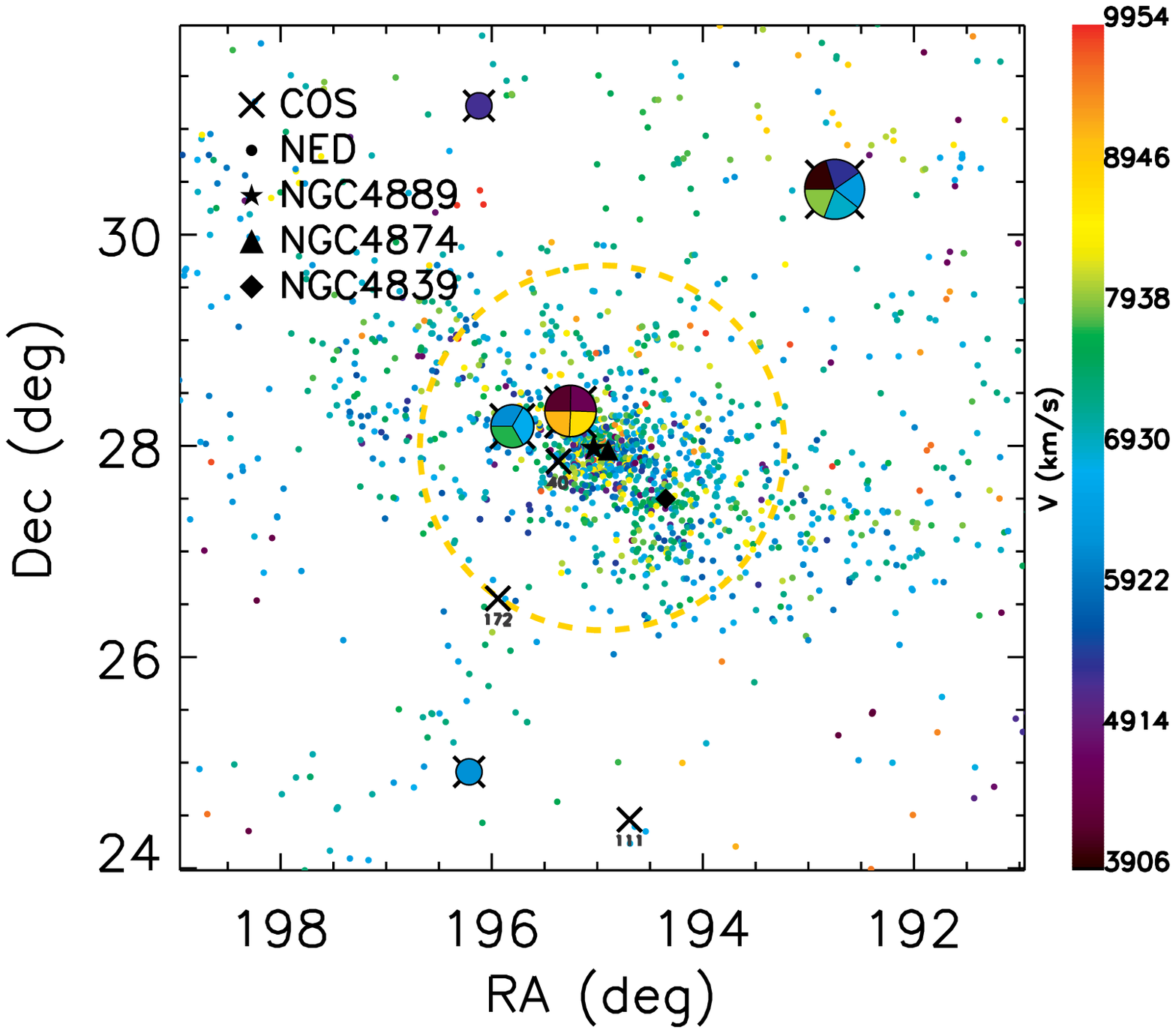}
  \includegraphics[width=1.2\columnwidth]{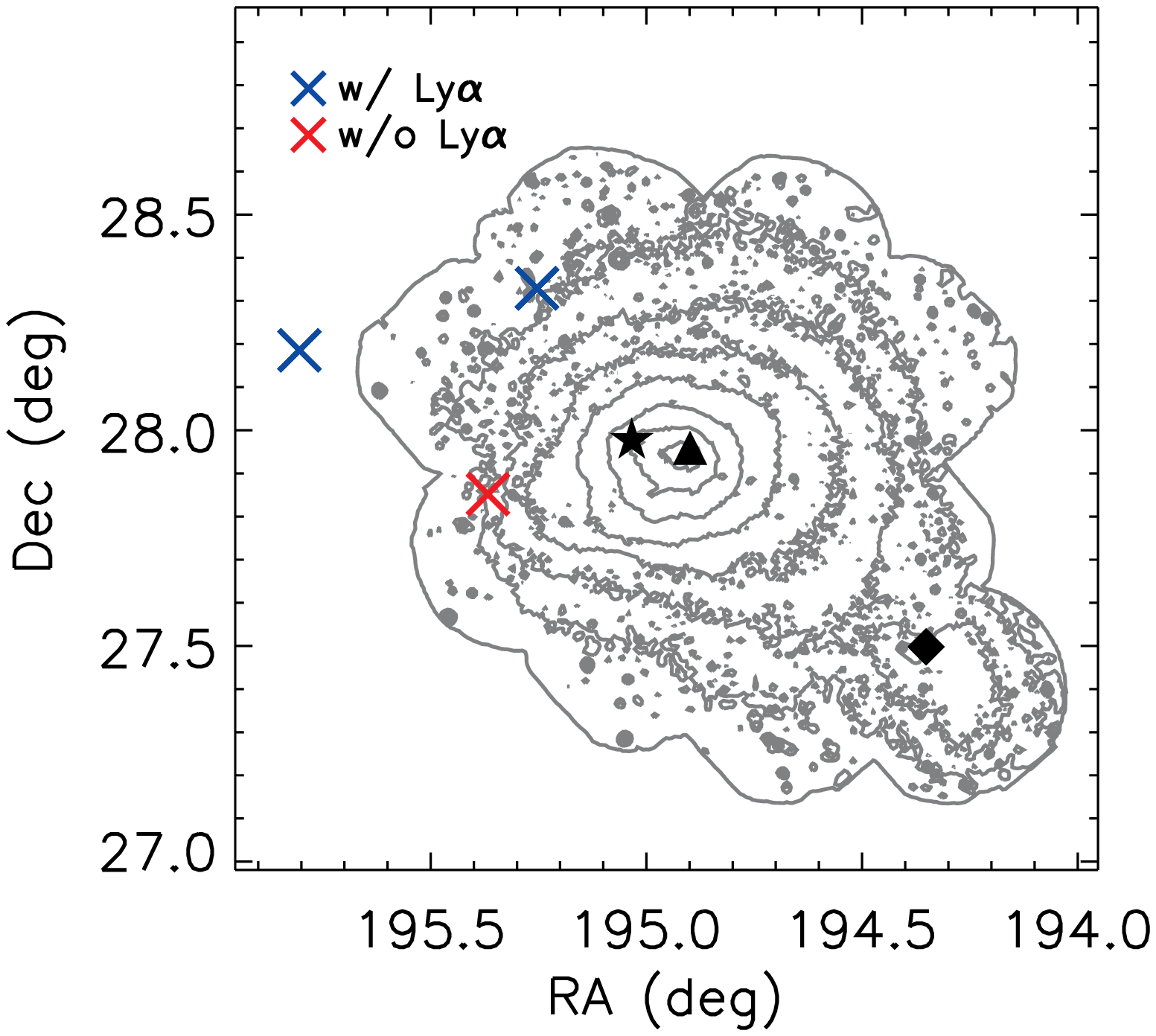}}
\caption{{\it Top}: COS sightlines (crosses) and the detected \lya absorbers (colored circles) in the Coma Cluster.  The color corresponds to the velocity of a \lya absorber as noted in the color bar to the right and if a sightline has multiple absorbers the circle has slices of different colors. Small dots are galaxies from NED with colors corresponding to their velocities. The yellow dashed circle represents \rvir\ for Coma (2.9 Mpc). 
The numbers next to crosses show detection limits (see Table~\ref{absorbers.tab}).  
{\it Bottom}: The 3 innermost sightlines are shown as crosses with the hot ICM from the XMM-Newton X-ray map \citep{Briel2001a} binned by $4\times4$ pixels in the background. The contour levels are 5, 10, 20, 40, 80, 130, 180 counts s$^{-1}$ deg$^{-2}$. The outermost solid line presents the field-of-view of the X-ray mosaic. In both panels the giant elliptical galaxies, NGC 4889 (star) at $v=6495\kms$, NGC 4874 (triangle) at $v=7176\kms$, and NGC 4839 (diamond) at $v=7362\kms$ are shown.}
\label{coma_plot.fig}
\end{center}
\end{figure*}

The target QSOs were selected to be UV bright and distributed between the center of the Coma Cluster and $\sim2 \times$ the virial radius (Figure~\ref{coma_plot.fig}).  Table~\ref{SLs.tab} shows a basic journal of the Coma Cluster observations. 
Three of the QSO sightlines pass through the central region of the cluster, within 0.5\rvir, one sightline is at approximately \rvir, and four sightlines are located at 1.27--2.04\rvir.
All sightlines were chosen by cross-matching the V{\'e}ron-Cetty QSO catalog 13th ed. \citep{Veron-Cetty2010a} with the GALEX database \citep{Martin2005}. The GALEX data provide an estimate of the FUV flux level, with the FUV imaging band covering the range $1344 - 1786\,\mbox{\AA}$ (effective wavelength 1516~\AA).  The SDSS images were checked to ensure there were no obvious stars or galaxies interfering with the GALEX flux.  All our targets were selected to be brighter than 18.75 AB mag to
reach the desired S/N level of 10 in 5 orbits or less (according to the COS Exposure Time Calculator).   
The pipeline processed data using {\it calcos} v2.21 were extracted from the MAST archive. For coaddition of the individual spectra, we employed the IDL routine, \texttt{coadd\_x1d.pro} version 3.3 \citep{Danforth2010a}.

In this paper we also include 10 COS updates to the warm gas map of the Virgo Cluster from Y12. Four sightlines were only observed with HST/STIS at the time of the Y12 publication (3C273, PG1216+069, RXJ1230.8+0115, and QSO-B1229+204) and now have HST/COS data, 5 additional HST/COS sightlines in the region were found in the HST archive, and 1 sightline (J1232+0603) is included from our HST/COS observations in Cycle 21 (PID: 13383). 
The data processing and spectrum coaddition were done in the same way as the Coma sightlines.

\begin{deluxetable*}{cccccrc}
\tabletypesize{\scriptsize}
\tablecaption{\label{SLs.tab} Summary of the COS Observations}
\tablewidth{0pt}
\tablehead{
  \colhead{QSO} &
  \colhead{Short Name} &
  \colhead{$R.A.$} &
  \colhead{$decl.$} &
  \colhead{$z_{\rm QSO}$} &
  \colhead{$t_{\rm exp}$} &
  \colhead{d$^{a}$} \\
  \colhead{} &
  \colhead{} &
  \colhead{(J2000,$^{\circ}$)} &
  \colhead{(J2000,$^{\circ}$)} &
  \colhead{} &
  \colhead{(s)} &
  \colhead{($R/R_{\rm vir}$)} }
\startdata
HB89 1259+281  & HB89 1259+281 & 195.3673 &   27.8518 & 0.24  &   13649            & 0.22 \\
HB89 1258+285 &  HB89 1258+285 & 195.2536 &   28.3291 & 1.36  &   5161   & 0.25 \\
Ton 0694 & Ton 0694 & 195.8043 &   28.1855 & 1.30  &   4822                  & 0.45 \\
RX J1303.7+2633 & J1303+2633 & 195.9416 &   26.5539 & 0.44  &   7016           & 0.97 \\
FBQS J1252+2913$^{b}$ & J1252+2913  &  193.1041 & 29.2226 & 0.82  &   6886           & 1.27 \\
Ton 0133 & Ton 0133 & 192.7513 &   30.4283 & 0.65  &   6589                   & 1.80 \\
FBQS J130451.4+245445 &  J1304+2454 & 196.2142 &   24.9128 & 0.60  &   5135     & 1.89 \\
SDSS J130429.04+311308.2 & J1304+3113 & 196.1210 &   31.2190 & 0.80  &   9232  & 1.96 \\
SDSS J125846.67+242739.1 & J1258+2427 & 194.6944 &   24.4609 & 0.37  &   7546  & 2.04 
\enddata
\tablecomments{$^a$Distance from the center of the cluster at $(R.A., decl.)=$(12:59:48.7, +27:58:50; Abell et al. 1989\nocite{Abell1989a}). $^{b}$ This sightline was wiped out by a LLS. }
\end{deluxetable*}

\begin{deluxetable*}{ccrrrcrlr}
\tabletypesize{\scriptsize}
\tablecaption{\label{absorbers.tab} COS sightlines with/without \lya and metal absorbers in the Coma Cluster}
\tablewidth{0pt}
\tablehead{
  \colhead{Name} &
  \colhead{Line} &
  \colhead{$\lambda_{\rm obs} $} &
  \colhead{$v$} &
  \colhead{$W$} &
  \colhead{$\sigma_{\rm W}$} &
  \colhead{$W_{\rm limit}$\tablenotemark{a}} &
  \colhead{log N \tablenotemark{b}} &
  \colhead{SL\tablenotemark{c}} \\
  \colhead{} &
  \colhead{} &
  \colhead{(\AA)} &
  \colhead{(\kms)} &
  \colhead{(\mA)} &
  \colhead{(\mA)} &
  \colhead{(\mA)} &
  \colhead{(cm$^{-2}$)} &
  \colhead{}
  }
\startdata
 FBQSJ1304+2454 &    HI1215 &   1240.96 & 6240 &  36 &   8 &  54 & 12.8 &  4.5 \\
  HB89 1258+285 &    HI1215 &   1232.69 & 4200 & 187 &  13 &  80 & 13.7 & 14.4 \\
                &    HI1215 &   1234.27 & 4590 & 215 &  15 &  80 & 13.8 & 14.3 \\
                &  SiII1190 &   1234.27 & 4590 & 163 &  17 &  80 & 13.6 &  9.6 \\
                &  SiII1193 &   1234.27 & 4590 &  98 &  12 &  80 & 13.3 &  8.2 \\
                &    HI1215 &   1251.17 & 8760 & 451 &  22 &  80 & 14.9 & 20.5 \\
                &  CII*1335 &   1251.17 & 8760 &  69 &  15 &  80 & 13.2 &  4.6 \\
                &    HI1215 &   1252.87 & 9180 & 198 &  14 &  80 & 13.7 & 14.1 \\
 SDSSJ1304+3113 &    HI1215 &   1237.07 & 5280 & 324 &  11 &  55 & 14.2 & 29.5 \\
        TON0694 &    HI1215 &   1240.83 & 6210 & 591 &  17 &  77 & 15.9 & 34.8 \\
                & SiIII1206 &   1240.83 & 6210 & 342 &  25 &  77 & 14.3 & 13.7 \\
                &   CII1334 &   1240.83 & 6210 & 145 &  29 &  77 & 13.5 &  5.0 \\
                &    HI1215 &   1243.02 & 6750 &  81 &  15 &  77 & 13.2 &  5.4 \\
                &    HI1215 &   1247.16 & 7770 & 126 &  16 &  77 & 13.5 &  7.9 \\
        TON0133 &    HI1215 &   1231.47 & 3900\tablenotemark{d} & 134 &  13 &  50 & 13.5 & 10.3 \\
        TON0133 &    HI1215 &   1231.84 & 3990 & 108 &  10 &  50 & 13.4 & 10.8 \\
                &    HI1215 &   1237.43 & 5370 & 316 &  14 &  50 & 14.2 & 22.6 \\
                &    HI1215 &   1241.56 & 6390 & 142 &  13 &  50 & 13.5 & 10.9 \\
                &    HI1215 &   1243.75 & 6930 & 465 &  12 &  50 & 15.0 & 38.8 \\
                &    HI1215 &   1248.01 & 7980 & 198 &  14 &  50 & 13.0 & 14.1 \\
  HB89 1259+281 &    HI1215 &   1258.70 & 0 &   0 &  0 &  40 & 12.9\tablenotemark{e} &  0.0 \\
   RXJ1303+2633 &    HI1215 &   1263.45 & 0 &   0 &  0 & 172 & 13.7\tablenotemark{e} &  0.0 \\
 SDSSJ1258+2427 &    HI1215 &   1280.83 & 0 &   0 &  0 & 111 & 13.4\tablenotemark{e} &  0.0
\enddata
\tablecomments{$^aW_{\rm limit}$ is estimated at 1244~\AA. $^b$b is assumed to be 30\kms. $^c{\rm SL}= W/\sigma_W$. $^d$This is lower than the $3\sigma$ velocity cut, $6930-3\times1008=3906$\kms, and is not included in the results throughout the paper. $^eW_{\rm limit}$ is converted to column density.} 
\end{deluxetable*}

\begin{deluxetable*}{crrrrrrlr}
\tabletypesize{\scriptsize}
\tablecaption{\label{virgo.tab} COS sightlines in the Virgo Cluster updated from Y12}
\tablewidth{0pt}
\tablehead{
  \colhead{Name} &
  \colhead{Line} &
  \colhead{$\lambda_{\rm obs}$} &
  \colhead{$v$} &
  \colhead{$W$} &
  \colhead{$\sigma_{\rm W}$} &
  \colhead{$W_{\rm limit}$} &
  \colhead{log N} &
  \colhead{SL} \\
  \colhead{} &
  \colhead{} &
  \colhead{(\AA)} &
  \colhead{(\kms)} &
  \colhead{(\mA)} &
  \colhead{(\mA)} &
  \colhead{(\mA)} &
  \colhead{(cm$^{-2}$)} 
  }
\startdata
J1232+0603\tablenotemark{c} & \HI & 1220.05 & 1080 & 187 & 17 & 63 & 13.7 & 11.0 \\
& \HI & 1225.27 & 2370 & 155 & 16 & 63 & 13.6 & 9.7 \\
3C273\tablenotemark{d} & \HI & 1219.80 & 1020 & 394 & 2 & 6 & 14.5 & 197.0 \\
& \SiII & 1219.80 & 1020 & 7 & 1 & 6 & 12.1 & 7.0 \\
& \SiIV & 1219.80 & 1020 & 13 & 1 & 6 & 12.4 & 13.0 \\
& \FeII & 1219.80 & 1020 & 33 & 2 & 6 & 12.8 & 16.5 \\
& \HI & 1222.11 & 1590 & 382 & 2 & 6 & 14.5 & 191.0 \\
& \SiII & 1222.11 & 1590 & 10 & 1 & 6 & 12.3 & 10.0 \\
& \SiII & 1222.11 & 1590 & 7 & 1 & 6 & 12.2 & 7.0 \\
& \CII & 1222.11 & 1590 & 12 & 1 & 6 & 12.4 & 12.0 \\
& \SiIII & 1222.11 & 1590 & 19 & 2 & 6 & 12.6 & 9.5 \\
& \HI & 1224.42 & 2160 & 34 & 2 & 6 & 12.8 & 17.0 \\
& \SiII & 1224.42 & 2160 & 11 & 1 & 6 & 12.3 & 11.0 \\
& \SiIV & 1224.42 & 2160 & 8 & 1 & 6 & 12.2 & 8.0 \\
& \HI & 1224.91 & 2280 & 44 & 2 & 6 & 12.9 & 22.0 \\
PG1216+069\tablenotemark{d} & \SiII & 1223.33 & 1890 & 49 & 5 & 25 & 13.0 & 9.8 \\
& \SiII & 1223.33 & 1890 & 69 & 3 & 25 & 13.2 & 23.0 \\
& \HI & 1223.33 & 1890 & 2239 & 16 & 25 & 19.0 & 139.9 \\
& \SiII & 1223.33 & 1890 & 162 & 6 & 25 & 13.6 & 27.0 \\
& \OI & 1223.33 & 1890 & 87 & 8 & 25 & 13.2 & 10.9 \\
& \SiII & 1223.33 & 1890 & 99 & 10 & 25 & 13.3 & 9.9 \\
& \CII & 1223.33 & 1890 & 119 & 6 & 25 & 13.4 & 19.8 \\
& \SiII & 1223.33 & 1890 & 57 & 8 & 25 & 13.1 & 7.1 \\
& \FeII & 1223.33 & 1890 & 23 & 5 & 25 & 12.6 & 4.6 \\
RXJ1230.8+0115\tablenotemark{d} & \HI & 1221.75 & 1500 & 174 & 3 & 9 & 13.7 & 58.0 \\
& \HI & 1222.60 & 1710 & 628 & 2 & 9 & 16.2 & 314.0 \\
& \SiII & 1222.60 & 1710 & 32 & 2 & 9 & 12.8 & 16.0 \\
& \SiIII & 1222.60 & 1710 & 186 & 4 & 9 & 13.7 & 46.5 \\
& \SiII & 1222.60 & 1710 & 57 & 2 & 9 & 13.1 & 28.5 \\
& \CII & 1222.60 & 1710 & 79 & 2 & 9 & 13.2 & 39.5 \\
& \SiIV & 1222.60 & 1710 & 71 & 2 & 9 & 13.2 & 35.5 \\
& \SiIV & 1222.60 & 1710 & 58 & 2 & 9 & 13.1 & 29.0 \\
& \CIV & 1222.60 & 1710 & 67 & 4 & 9 & 13.1 & 16.8 \\
& \CIV & 1222.60 & 1710 & 47 & 4 & 9 & 12.0 & 11.8 \\
& \AlII & 1222.60 & 1710 & 19 & 3 & 9 & 12.6 & 6.3 \\
& \HI & 1223.21 & 1860 & 83 & 3 & 9 & 13.2 & 27.7 \\
& \HI & 1225.03 & 2310 & 345 & 2 & 9 & 14.3 & 172.5 \\
QSO-B1229+204\tablenotemark{d} & \HI & 1220.41 & 1170 & 339 & 12 & 29 & 14.3 & 28.2 \\
& \HI & 1223.33 & 1890 & 289 & 12 & 29 & 14.1 & 24.1 \\
& \HI & 1226.00 & 2550 & 248 & 11 & 29 & 13.9 & 22.5 \\
VV2006-J125124+055420 & \HI & 1215.67 & 0 & 0 & 0 & 84 &  13.3\tablenotemark{b} & 0.0 \\
VV2006-J130524+035731 & \HI & 1218.35 & 660\tablenotemark{a} & 299 & 28 & 48 & 14.1 & 10.7 \\
SDSSJ123304-003134 & \HI & 1220.29 & 1140 & 368 & 11 & 45 & 14.4 & 33.5 \\
& \NV & 1220.29 & 1140 & 185 & 9 & 45 & 13.7 & 20.6 \\
& \SiIV & 1220.29 & 1140 & 28 & 6 & 45 & 12.7 & 4.7 \\
& \CIV & 1220.29 & 1140 & 82 & 7 & 45 & 13.2 & 11.7 \\
& \CIV & 1220.29 & 1140 & 136 & 13 & 45 & 13.5 & 10.5 \\
& \HI & 1225.40 & 2400 & 112 & 8 & 45 & 13.4 & 14.0 \\
& \HI & 1227.34 & 2880\tablenotemark{e} & 107 & 9 & 45 & 13.4 & 11.9 \\
SDSSJ115758-002220 & \HI & 1227.71 & 2970\tablenotemark{e} & 166 & 17 & 62 & 13.6 & 9.8 \\
IRAS11598-0112 & \HI & 1221.87 & 1530 & 199 & 26 & 167 & 13.7 & 7.7
\enddata
\tablecomments{This table is the same as Table~\ref{absorbers.tab} but for updates to the Virgo sightlines presented in Y12 (see this paper for the other lines). $^a$This is a lower velocity than 700\kms. $^b$W$_{\rm limit}$ is converted to column density. $^c$This sightline is from a new observation (PID 13383; PI Putman). $^d$These sightlines were found in the HST/COS archive and replace the STIS sightlines in Y12. $^e$These absorbers are beyond $3\sigma_{virgo}=2770$ \kms, but within the 3000 \kms~cut used in Y12.}
\end{deluxetable*}

\begin{figure*}[ht]
  \includegraphics[width=2.1\columnwidth]{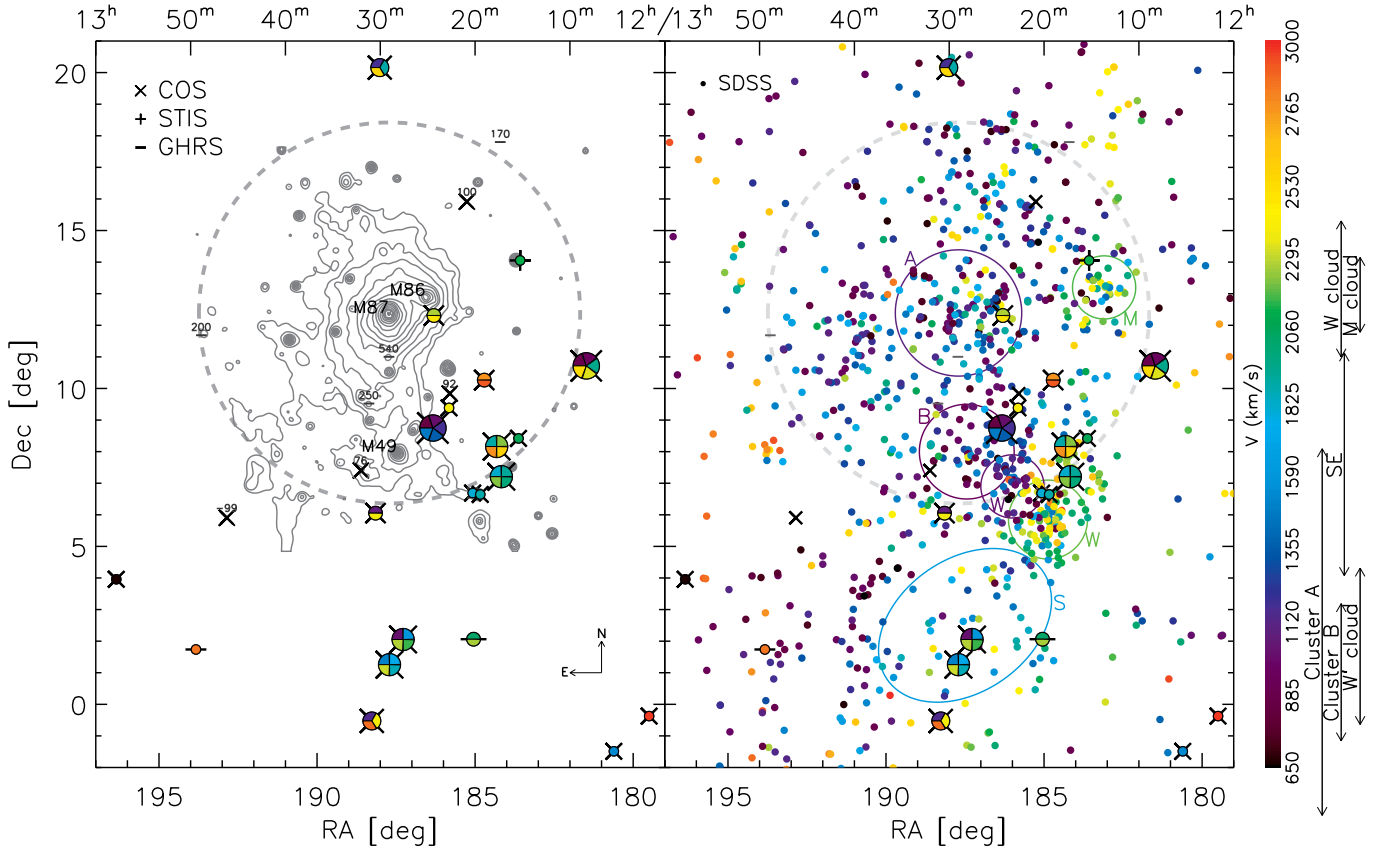}
\caption{An updated \lya absorber map for the Virgo Cluster that includes the data in Table~\ref{virgo.tab} and Y12. The virial radius of the cluster is noted with the large grey dashed circle (1.57 Mpc) and the sightlines are marked according to the instruments (cross - COS, plus - STIS, bar - GHRS).  If there is no detection the limit in \mA~is noted above the symbol.  A circle over the symbol represents the detection of one or more Ly$\alpha$ absorbers with the colors representing the velocities of the absorbers (velocity color bar is to the right).   {\it Left}: The ROSAT X-ray map in the background with the level of the contours corresponding to that of \citet{Bohringer1994a} except that the lowest level here is 5$\sigma$.  
{\it Right}: The SDSS galaxies color-coded by their radial velocities in the background with the substructures, A (Cluster A), B (Cluster B), M (M cloud), W (W cloud), \Wprime (\Wprime cloud) and S (Southern Extension) noted \citep[Y12,][]{binggeli87}. The color of each substructure circle/ellipsoid corresponds to its mean velocity. The velocity ranges of each substructure are also noted by  arrows on the right side of the color bar.}
\label{virgo_plot.fig}
\end{figure*}

\section{Results}

 We detect 14 \lya absorbers towards 5 of the 8 sightlines within 2 \rvir~ of Coma in the velocity range $v_{\rm coma}\pm3\sigma_{\rm v}$ (6 absorbers in this area within $v_{\rm coma}\pm1\sigma_{\rm v}$, where $v_{\rm coma}=6930\kms$ and $\sigma_{\rm v}$=1008\kms \citep{Struble1999a}).  The absorbers are shown in Figure~\ref{coma_plot.fig} and the detected lines and limits are listed in Table~\ref{absorbers.tab}.  Table~\ref{absorbers.tab} also includes one absorber at 3900~\kms~that is not included in the figures and analysis because it is just beyond the $3\sigma_{\rm v}$ limit at 3906 \kms.  We note in the text if the inclusion of this absorber would change any results. Two of the sightlines without an absorber have W$_{\rm limit}$ values significantly higher than the sightlines with detections, while one of the non-detection sightlines (the innermost one) has the lowest W$_{\rm limit}$ of the sample.
 The column densities of the \lya absorbers range from $10^{12.8}\cm-2$ to $10^{15.9}\cm-2$, with a mean value of $10^{13.9}\cm-2$ and a median of $10^{13.7}\cm-2$. Three of the \lya absorbers show corresponding metal absorption lines.  All detections have a significance level (${\rm SL}= W/\sigma_W$) $\geq 4.5$. 

For the Virgo Cluster, towards the 10 updated sightlines (see \S 2), we identify 19 \lya absorbers in the Virgo velocity range used in Y12
($700 - 3000$\kms) and 1 additional \lya absorber at 660\kms (see Table~\ref{virgo.tab}).
Combined with the findings of Y12,  this results in 45 \lya absorbers towards 22 sightlines observed by HST/COS with column densities ranging from $10^{12.0}\cm-2$ to $10^{19.0} \cm-2$.  
The results are shown in Figure~\ref{virgo_plot.fig} and include six remaining STIS or GHRS sightlines, resulting in a total of 49 \lya absorbers in the vicinity of the Virgo Cluster.  The mean column density of all 49 absorbers is $10^{17.2}\cm-2$ and the median is $10^{13.3}\cm-2$.  
While no metal lines were found associated with the \lya absorbers in Y12, 6 of the \lya absorbers from the updated sightlines have metal detections. All of the new sightlines with detected metal lines are deeper than the observations presented in Y12.  The central velocity and dispersion of Virgo varies depending on the galaxies included.  We use  $v_{\rm Virgo}=1138\kms$ and $\sigma_{\rm v}$=544\kms~as found by \cite{Mei2007a}.
The velocity range used in Y12 was $700 - 3000$\kms~due to the Milky Way damping wings at the low velocity end, and Virgo's infalling substructures that extend out to 3000\kms~at the high velocity end.  We present absorbers over this velocity range in the table and figures and note if the inclusion of the two absorbers beyond $v_{\rm virgo}\pm3\sigma_{\rm v} = 2770$\kms~ affect any results.

\subsection{The Distribution of \lya Absorbing Gas}

We investigate the distribution of the \lya absorbers in comparison with the known structure of the Coma Cluster.
In the top panel of Figure~\ref{coma_plot.fig}, the distribution of our sightlines is shown in relation to the galaxies in and around the Coma Cluster.  There are 4 sightlines within \rvir~of the Coma Cluster (see also Table~\ref{SLs.tab}) and 2 of these sightlines have \lya absorbers.  The bottom panel of Figure~\ref{coma_plot.fig} zooms in on the three sightlines closest to center of the cluster with the XMM-Newton x-ray map in the background \citep{Briel2001a}. The innermost sightline without absorbers shown in both panels, HB89 1259+281, has the lowest detection limit of our sightlines (40 \mA).  The other two inner sightlines shown in both panels have absorbers. HB89 1258+285, at 0.25 \rvir, has 4 \lya absorbers, but their velocities are all outside of 1\sigmav\ for the Coma Cluster. Ton 0694, at 0.45 \rvir, has 3 \lya absorbers and their velocities are all within 1\sigmav (see \S3.2).
 The sightline just within the virial radius only shown in the top panel (J1303+2633) has no absorbers but also an unusually high detection limit (172 \mA).
For the 4 sightlines at $>$ \rvir, we detect 7 \lya absorbers within 3\sigmav\ and 3 \lya absorbers within 1\sigmav\ towards 3 of the sightlines.

The updates to the Virgo sightlines are all beyond \rvir~ and include sightlines out to 2.58\rvir~to track the large scale flows further from the center of the cluster.  The distribution of absorbers relative to the Virgo Cluster does not substantially changed with these additions.   The main difference is there are now two non-detections beyond \rvir~in regions of lower galaxy density.
Both the Coma and Virgo clusters have absorbers at maximal separation in the fields shown in Figures~\ref{coma_plot.fig} and ~\ref{virgo_plot.fig} with a velocity separation of only $\sim150$\kms.

The absorbers can be compared to the other baryonic components of the clusters.
The relationship of the absorbers to the galaxies is discussed in \S\ref{galsect}. In the context of the relation to the detected hot x-ray gas, for Coma the only sightline within the outermost contour of the X-ray map (5 counts s$^{-1}$ deg$^{-2}$) is the clear non-detection (HB89 1259+281, the sightline with the deepest limit).  The sightline at the outer edge of these contours (HB89 1258+285) has absorbers, but they are likely to be in the foreground/background given their velocities relative to Coma.  Thus, there is no indication of warm gas within the detected hot gas of the Coma Cluster.  The newly added sightlines for the Virgo Cluster are located outside of \rvir\ and do not change the conclusions in Y12 that the warm absorbers also avoid the detected hot gas.

\begin{figure}[]
\begin{center}
\includegraphics[width=1.05\columnwidth]{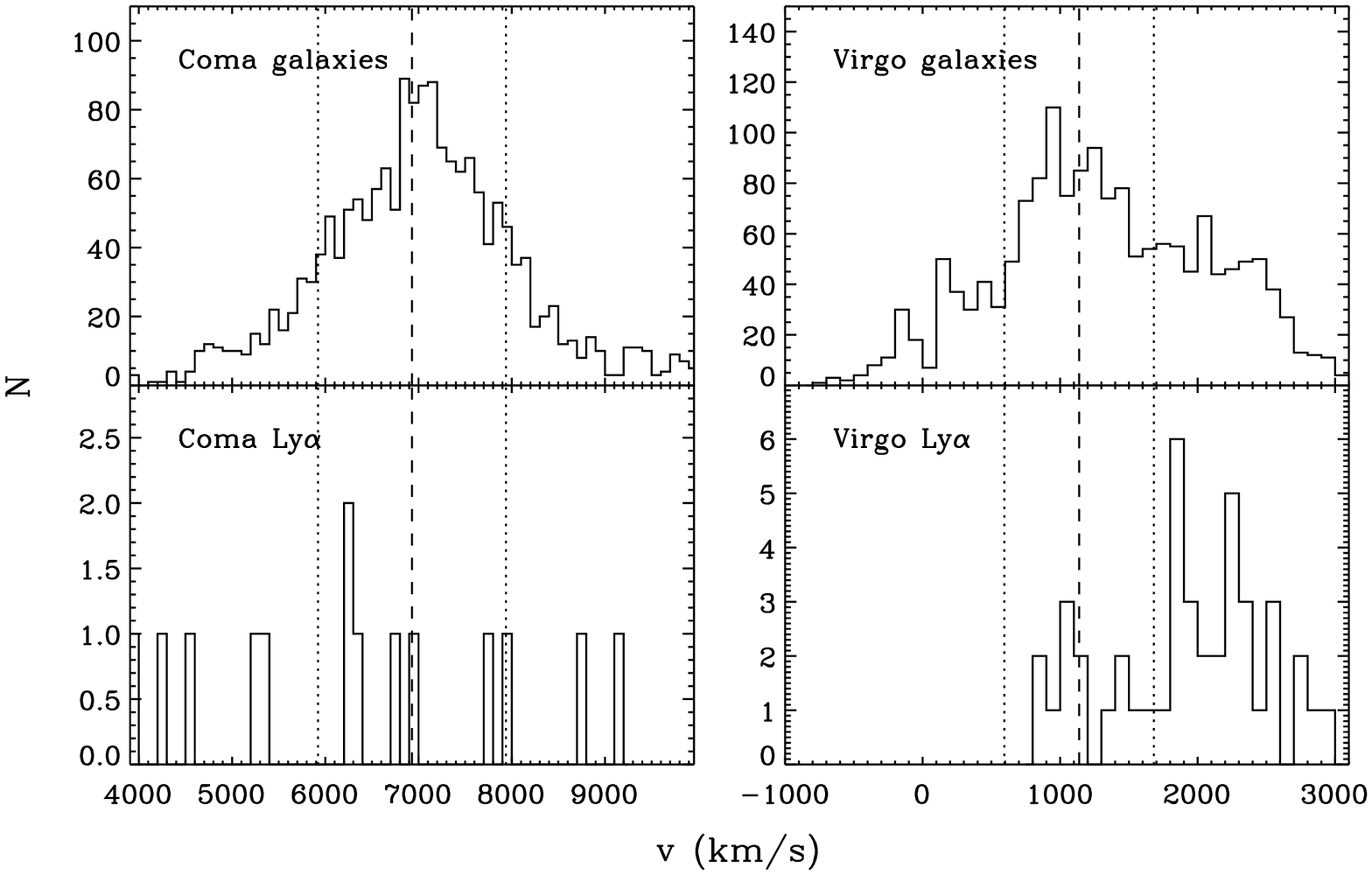}
\caption{Velocity distribution of galaxies (top) and \lya absorbers (bottom) for the Coma Cluster (left) and the Virgo Cluster (right). Dashed lines and dotted lines represent the central velocity and velocity dispersion for each cluster. }
\label{vdist.fig}
\end{center}
\end{figure}

The velocity distributions of Coma's 14 \lya absorbers and the galaxies shown in Figure~\ref{coma_plot.fig} are presented on the left of Figure~\ref{vdist.fig}.  With the current sample, the absorbers extend evenly across the $3\sigma_{\rm v}$ velocity range shown and this is in contrast to the galaxy velocities in this same area of sky.   The dispersion of the Coma galaxy velocities is 1031~$+/- 19$ \kms~with a mean velocity at 7020 \kms, while the absorbers have a dispersion of 1626~$+/- 236$ \kms, with a mean at 6403\footnote{The inclusion of the absorber at 3900 \kms~would change these values to a dispersion of 1695\kms~and a mean velocity of 6236 \kms.} \kms.  The errors on the dispersion values were obtained with a bootstrap analysis.
Therefore the Coma absorbers show a significantly larger dispersion than the galaxies, with a shift towards a somewhat lower mean velocity.  We can only sample one side of the Virgo velocity range, but the absorbers seem to peak at 1-2\sigmav~beyond the central velocity of the cluster galaxies (right panels of Figure~\ref{vdist.fig}).  For both clusters the ratio of absorbers to galaxies is higher beyond $\sigma_{\rm v}$ than within $\sigma_{\rm v}$.  Dispersion values for the absorbers we detect in and around Virgo are not available given the cutoff in our observable velocity range.  If one examines the dispersion at only v$_{\rm Virgo} + 3\sigma_{\rm v}$ it is higher for the absorbers than galaxies shown in Figure~\ref{virgo_plot.fig} within \rvir~and lower outside of \rvir.

\begin{figure}[]
\begin{center}
  \includegraphics[width=1\columnwidth]{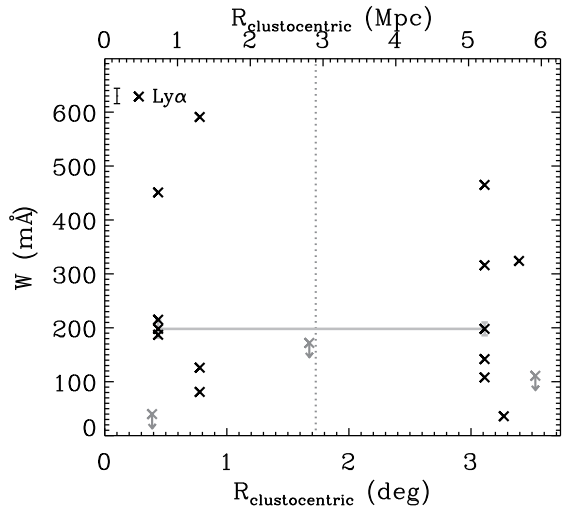}
\caption{Cluster impact parameter vs. EW of Coma's \lya absorbers. Gray data points with downward arrows indicate the detection limits for the sightlines without \lya absorbers. The vertical dotted line presents \rvir\ and the gray line connects the median EW in two bins divided by \rvir. The mean error is marked on the top left. }
\label{Rclt_W.fig}
\end{center}
\end{figure}

The absorption line strength as a function of cluster impact parameter is shown for Coma in Figure~\ref{Rclt_W.fig}. There is no signature of a relationship with impact parameter for the 14 absorbers in our sample.
For Virgo the results do not change from Y12; i.e., we do not see a trend unless the infalling substructures are taken out, and then there is a hint of increasing column density with increasing impact parameter.

\subsection{\lya Absorbers in Relation to Galaxies}
\label{galsect}

The galaxy distribution of the Coma Cluster is elongated in the north-east and south-west directions in Figure~\ref{coma_plot.fig}. The galaxies shown cover the velocity range $v_{\rm Coma}\pm3\sigma_{\rm v}$ and
are from the NASA/IPAC Extragalactic Database\footnote{http://ned.ipac.caltech.edu/} with their photometry cross-matched with the SDSS DR12 photometric catalog \citep{Alam2015a}. The elongation of the galaxy distribution is linked to large scale filamentary structure. In position-velocity space the galaxy distribution has distinct groups centered on NGC 4874 and NGC 4839 (labeled in Figure~\ref{coma_plot.fig}), and the NGC 4839 group is falling onto the cluster \citep{Colless1996a, Neumann2001a}.
We do not have sightlines along the elongated galaxy distribution, so we cannot investigate a bias for absorbers to follow this structure.  Below, we focus our analysis on the individual galaxies around each sightline.

\begin{figure}[]
\begin{center}
  \includegraphics[width=1\columnwidth]{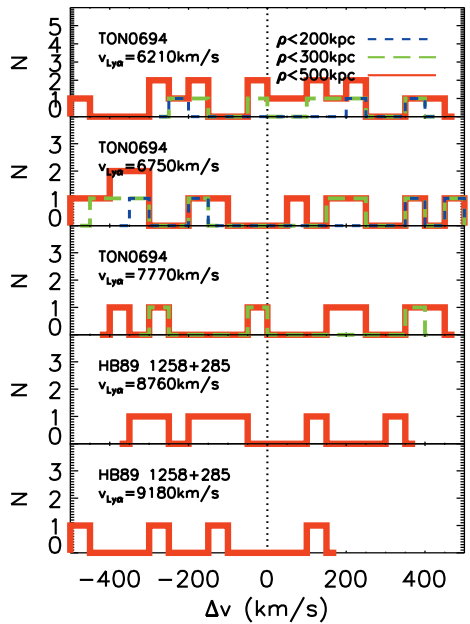}
\caption{The Coma \lya absorbers with galaxies (from NED) within 500\kms\ and 200 (short dashed blue), 300 (long dashed green), and 500 (solid red) kpc are shown by each panel.  The name of the sightline and the velocity of each \lya absorber is noted in the upper left of each panel.  The top panel has a different y-axis to include the legend.}
\label{v_gal.fig}
\end{center}
\end{figure}

In our investigation of the connection between galaxies and \lya absorbers in the Coma Cluster, we examine all galaxies within $\Delta v < 500$\kms\ and $\rho < 200,~300,~500$ kpc of absorbers as shown in Figure~\ref{v_gal.fig}. 
Of our 14 \lya absorbers, only 2 towards TON 0694 (\rvir = 0.45) have a galaxy within 200 kpc and the third \lya absorber along this sightline has a galaxy within 300 kpc.  All three of these absorbers have the closest galaxy within a velocity difference of 300 \kms~and it seems likely they represent associated galaxies and absorbers.  The absorber with the most nearby galaxies at 6210 \kms also has associated metal lines.  These nearby galaxies tend to be fainter than typical Coma galaxies and belong to the red sequence.  There are no other galaxies within 300 kpc and 500 \kms\ of an absorber in our sample.  If we extend the impact parameter to 500 kpc, two of the \lya absorbers towards HB89 1258+285 have galaxies within 500~\kms, but these are the same galaxies found to be closer to the TON 0694 absorbers.

All other Coma \lya absorbers (9) show no galaxies within 500 kpc and 500\kms. These absorbers are also either beyond the virial radius of Coma, or beyond 1\sigmav~ along the line of sight.   Specifically, the outer sightlines beyond \rvir\ have absorbers within 1\sigmav, and the 3 absorbers within \rvir\ that do not have associated galaxies are beyond 2\sigmav.
Though not all small galaxies are recovered in this comparison, the results imply that the majority of \lya absorbers detected in Coma arise from the intergalactic medium surrounding the Coma Cluster.  This is consistent with the low column densities of most absorbers.

For the Virgo Cluster, when adopting the galaxy sample from NED with no magnitude cuts\footnote{The galaxies used for the Virgo Cluster figures are from SDSS with r$_{mag} < 17.77$ and \cite{Yoon2013a} used these SDSS galaxies together with NED galaxies with a similar magnitude cut.}, many \lya absorbers can be found within 300~\kms~and 300 kpc of galaxies. However, since the Virgo Cluster is at 16.5 Mpc and Coma is at 105 Mpc, if we adopt the same absolute magnitude limit as Coma there are substantially less associations.  It is unclear if this means significantly more galaxy-absorber associations would be found for Coma if fainter galaxies were included.  The Virgo sightlines clearly trace large dark matter sub-structures (see right panel of Figure~\ref{virgo_plot.fig}), and the line between what is CGM and IGM is likely to be blurrier in these regions.  Overall, the results remain consistent with Y12 that the bulk of the warm gas detected is related to inflowing IGM beyond the cluster's virial radius; there is only one sightline within \rvir~that has absorbers within 1\sigmav~of Virgo.  The new result with the updated sightlines is the detection of multiple metal lines.   Using the NED galaxy sample, all but one of the \lya absorbers with associated metal lines (along the 3C273 sightline) have a galaxy within 300 kpc and 300 \kms.

For Virgo we were able to compare the absorbers to blind HI emission surveys in Y12.  Coma does not have a blind HI survey of the entire region shown in Figure~\ref{coma_plot.fig}, but there are HI observations of 19 galaxies within \rvir\ (in projection) of the Coma Cluster \citep{Bravo-Alfaro2000a}.   
The two inner sightlines, HB89 1258+285 (at 0.25 R$_{vir}$) and HB89 1259+281 (at 0.22 R$_{vir}$), are closest to \HI\ galaxies mapped by \cite{Bravo-Alfaro2000a}.  HB89 1258+285 has two \lya absorbers at 8760\kms~and 9180\kms~that are close in position and velocity to \HI galaxies, i.e., one galaxy at $\rho = 375$ kpc and a velocity of 8884\kms, one at $\rho = 365$ kpc and 8426\kms, and one at $\rho =  323$ kpc and 7758\kms. 
The impact parameters are $>300$ kpc, so the absorbers are not within the halos of these galaxies, but the \lya absorbers may be part of a larger gaseous structure that includes the galaxies.  Given the higher velocities, this structure is most likely infalling to the cluster.  Unfortunately, we do not have direct distance determinations to further constrain their location.

The other sightline near \HI\ galaxies, HB89 1259+281, does not have absorbers to deep limits.  This is despite the fact that the \HI galaxy NGC 4921 has an impact parameter of only 64 kpc.  When we examine this galaxy with new 21cm \HI data from the WSRT Coma mosaic (Paolo Serra, private communication), the extent of the \HI detection ends at $\sim 40$ kpc at a column density $\sim 10^{19}$\cm-2.  NGC 4921 is at 5479\kms\ (within 1.5$\sigma_{\rm v{\rm Coma}}$) and its gas may have been truncated by the hot ICM shown in Figure~\ref{coma_plot.fig}.  Another example of the non-detection of \lya absorption in close proximity to HI emission is the galaxy pair NGC~4532/DDO~137 just beyond \rvir~and within $2\sigma_{\rm v}$ of the Virgo Cluster \citep{pearson16}.   These types of non-detections are interesting in terms of galaxy quenching and are worth further investigation in the future.

\begin{deluxetable*}{cccccc}
\tabletypesize{\scriptsize}
\tablecaption{\label{fcover_coma.tab} Covering fraction of \lya absorbers for the Coma Cluster}
\tablewidth{0pt}
\tablehead{
  \multicolumn{2}{c}{} &
  \multicolumn{2}{c}{$|v-v_{\rm Coma}|<1\sigma_{\rm v}$} &
  \multicolumn{2}{c}{$|v-v_{\rm Coma}|<2\sigma_{\rm v}$} \\
  \colhead{$W$ } &
  \colhead{log \nhi } &
  \colhead{$r\le$ \rvir} &
  \colhead{$r>$ \rvir} &
  \colhead{$r\le$ \rvir} &
  \colhead{$r>$ \rvir}\\
  \colhead{(\mA)} &
  \colhead{(cm$^{-2}$)} &
  \colhead{} &
  \colhead{} &
  \colhead{} &
  \colhead{} 
  }
\startdata
$>100$ & 13.3 & 0.33$_{-0.26}^{+0.17}$ (1/3) & 0.33$_{-0.26}^{+0.17}$ (1/3) & 0.67$_{-0.25}^{+0.16}$ (2/3) & 0.67$_{-0.25}^{+0.16}$ (2/3) \\
$>200$ & 13.8 & 0.25$_{-0.13}^{+0.24}$ (1/4) & 0.25$_{-0.13}^{+0.24}$ (1/4) & 0.50$_{-0.20}^{+0.20}$ (2/4) & 0.50$_{-0.20}^{+0.20}$ (2/4) \\
$>300$ & 14.1 & 0.25$_{-0.13}^{+0.24}$ (1/4) & 0.25$_{-0.13}^{+0.24}$ (1/4) & 0.50$_{-0.20}^{+0.20}$ (2/4) & 0.50$_{-0.20}^{+0.20}$ (2/4)
\enddata
\tablecomments{The covering fraction is  $(N_{\rm sightlines}$ with detections) / (Total $N_{\rm sightlines}$).  The error is estimated based on a likelihood analysis described in \cite{Chen2010a}. The problem is formulated in such way that if the gas covering fraction is k, then the likelihood of observing an ensemble of (n+m) sightlines, n of which show detections while m show no detections, is $prob = k^n \times (1-k)^m$ where $k$ can be solved by requiring maximum probability.
}
\end{deluxetable*}

\begin{deluxetable*}{cccccc}
\tabletypesize{\scriptsize}
\tablecaption{\label{fcover_virgo.tab} Covering fraction of \lya absorbers for the Virgo Cluster}
\tablewidth{0pt}
\tablehead{
  \multicolumn{2}{c}{} &
  \multicolumn{2}{c}{$v_{\rm Virgo}<v<v_{\rm Virgo}+1\sigma_{\rm v}$} &
  \multicolumn{2}{c}{$v_{\rm Virgo}<v<v_{\rm Virgo}+2\sigma_{\rm v}$} \\
  \colhead{$W$ } &
  \colhead{log \nhi } &
  \colhead{$r\le$ \rvir} &
  \colhead{$r>$ \rvir} &
  \colhead{$r\le$ \rvir} &
  \colhead{$r>$ \rvir}\\
  \colhead{(\mA)} &
  \colhead{(cm$^{-2}$)} &
  \colhead{} &
  \colhead{} &
  \colhead{} &
  \colhead{} 
  }
\startdata
$>100$ & 13.3 & 0.10$_{-0.05}^{+0.14}$ (1/10) & 0.38$_{-0.12}^{+0.14}$ (5/13) & 0.40$_{-0.13}^{+0.16}$ (4/10) & 0.69$_{-0.13}^{+0.10}$ (9/13) \\
$>200$ & 13.8 & 0.09$_{-0.05}^{+0.13}$ (1/11) & 0.29$_{-0.10}^{+0.13}$ (4/14) & 0.27$_{-0.12}^{+0.15}$ (4/11) & 0.43$_{-0.12}^{+0.13}$ (6/14) \\
$>300$ & 14.1 & 0.00$_{-0.00}^{+0.08}$ (0/12) & 0.25$_{-0.09}^{+0.12}$ (4/16) & 0.08$_{-0.04}^{+0.12}$ (1/12) & 0.44$_{-0.11}^{+0.12}$ (7/16) 
\enddata
\tablecomments{The relevant notes can be found in the Coma covering fraction table.
}
\end{deluxetable*}

\subsection{Covering Fraction}

The results for the covering fraction of \lya absorbers in the Coma Cluster are shown in Table~\ref{fcover_coma.tab}. 
The covering fraction of \lya absorbers within 1\sigmav\ or 2\sigmav\ when they are within \rvir\ or outside \rvir\ is computed with different sensitivity limits. 
The covering fraction at $|v-v_{\rm Coma}|<1\sigma_{\rm v}$ ($|v-v_{\rm Coma}|<2\sigma_{\rm v}$) within \rvir\ is 33\% (67\%) and it is 33\% (67\%) outside \rvir\  when log~\nhi~$>13.3$. 
 This trend does not change significantly at higher column densities and the covering fraction is the same for the absorbers within $2\sigma_{\rm v}$ and $3\sigma_{\rm v}$.  Therefore with our limited number of sightlines, we do not find differences in the covering fraction at different projected locations for the Coma Cluster. 

For comparison, the covering fraction of \lya absorbers in the Virgo Cluster is computed using the same criteria in Table~\ref{fcover_virgo.tab}. Since the detection of \lya absorbers below 1000 \kms\ is incomplete due to the Milky Way damping wings, we can only consider velocities where $v > v_{\rm Virgo}$ in the covering fraction. Despite the missing velocity range, for Virgo outside of \rvir\ for \nhi~$>10^{13.3}$~\cm-2 38\% of the sky is covered within $1\sigma_{\rm v}$ and 69\% within $2\sigma_{\rm v}$.  In contrast, the covering fraction within Virgo's \rvir\ for gas at log~\nhi~$>13.3$ within $1\sigma_{\rm v}$ ($2\sigma_{\rm v}$) is 10\% (40\%).  Therefore, though only half of the velocity range is being covered, thus far Virgo has a similar covering fraction to Coma in the outer regions and lower values than Coma within \rvir.  Virgo's difference in covering fraction inside and outside \rvir\ also holds at higher column densities, log~\nhi~$>$~13.8, and 14.1. If the velocity range is extended to $3\sigma_{\rm v}$ the covering fractions are similar to $2\sigma_{\rm v}$ with 50\% within \rvir\ and 77\% outside \rvir\ at log~\nhi$~>13.3$.

\section{Discussion}

This paper presents the distribution of warm gas, as traced by \lya absorbers, surrounding the Coma galaxy cluster.  This is only the second study of its kind for a cluster and we include an update to the first study of the Virgo Cluster presented in Y12.  We find the warm gas surrounds both clusters, 
and the absorbers' column densities and spatial-kinematic distributions indicate they largely represent the infall of the surrounding IGM.  In \S4.1 we discuss the absorbers in the context of the warm gas distribution for different mass dark matter halos.  In \S4.2 we discuss the relation of the absorbers to galaxies and in \S4.3 we examine the constraints the absorbers put on turbulent gas motions in the outskirts of clusters.   Finally, in \S4.4 we compare the results to recent simulations of a Coma-like and Virgo-like cluster.

\subsection{\lya Absorbers and the Most Massive Halos}

The warm IGM is predicted to flow into dark matter structures along filaments and go through a series of transitions as it encounters other gas and reaches faster speeds \citep[e.g.,][EBP15]{kravtsov12,joung12}.  At the intermediate mass scale of galaxies (\lessapprox{10$^{13}$~\msun}), large amounts of warm and warm-hot gas have been detected within their virial radii \citep{werk14,Chen2001a,Tumlinson2013a}. At the largest mass scale of galaxy clusters observed here ($\sim 10^{14-15}$ \msun), the majority of the warm gas is expected to be shock-heated to temperatures $>10^{7}$ K within the virial radius \citep{voit05,kravtsov12}.  We therefore expect to detect the warm gas flowing into the clusters, but not within the clusters themselves.   In particular, the gas flowing into clusters along cosmic filaments is expected to contain clumps of cooler gas \citep[][EBP15]{nagai07,Nagai2011} that may be analogous to the \lya absorbers.

The absorber results for the Virgo and Coma Clusters are largely consistent with the expectations from cosmological simulations (see \S4.4 for more details).
For the Virgo Cluster, we confirm the results of Y12 that the \lya absorbers are largely found in the outskirts in position-velocity space and avoid the region within the virial radius where the hot ICM exists.  All but 3 of the absorbers within the virial radius in projection (see Figure~\ref{virgo_plot.fig}) are beyond $1\sigma_{\rm v,Virgo}$, and are most likely associated with infalling substructures.
The 3 absorbers (along 1 sightline) within $1\sigma_{\rm v,Virgo}$ have numerous galaxies within 500 kpc and 500 \kms~ and are likely to represent stripped material from infalling galaxies.
For the Coma Cluster, the results also indicate the warm IGM is no longer found in abundance within the virial radius and avoids the hot ICM.   For the two sightlines that show absorption in projection within the virial radius in Figure~\ref{coma_plot.fig}, one of them has the four absorbers well beyond 
$1\sigma_{\rm v,Coma}$ and are likely associated with foreground/background infalling IGM, and the other has the three absorbers associated with galaxy halos (see \S4.3) and therefore the accretion of galaxies.

These cluster results stand in contrast to the observational results for the smaller dark matter halos of galaxies. Galaxies have a distinct relationship of increasing \lya column density with decreasing impact parameter \citep[e.g.,][]{Chen2001a,Prochaska2011a, Stocke2013a,Tumlinson2013a}.  This can be interpreted as part of the cooling of the halo gas as it approaches the central regions of the galaxy.  For the clusters we do not observe a relationship between impact parameter and column density (Figure~\ref{Rclt_W.fig}). As mentioned above, this is expected with the shock heating of the IGM in a cluster-scale massive halo within the virial radius. Future observations can investigate the column density-impact parameter relationship further and potentially determine the mass scale where this relationship ceases to hold.    

Another contrast point between the cluster-scale halos and galaxies is the covering fraction.
The covering fraction of \lya absorbers for galaxies approaches 100\% within the virial radius and generally drops to $<60$\% beyond the virial radius \citep[e.g.][]{Chen2001a,Tumlinson2013a,wakker09,Prochaska2011a,burchett16}.  These types of covering fractions are obtained by including absorbers within a wide velocity range; generally +/- 300-400 \kms~from the galaxy's systemic velocity, independent of galaxy type.   While the covering fraction inside \rvir~is lower for clusters than galaxies ($<70$\% within $2\sigma_{\rm v}$), the covering fraction beyond \rvir\ of the clusters is high in comparison to galaxies ($>60$\% within $2\sigma_{\rm v}$).   This difference may represent the shock heating of the gas within the virial radius for the clusters and larger coherent flows beyond the cluster virial radii in contrast to the narrow filaments that accrete onto galaxies.  For instance, \cite{joung12} found that 70\% of the mass influx for an L$_*$ galaxy is concentrated in only 17\% of the surface area. Compared to cluster simulations, however, the observed cluster covering fraction is high (see \S4.4).  
The larger covering fraction at $>$\rvir~for Virgo compared to Coma may be explained by the large number of infalling structures in more dynamically active systems and potentially less shock heating for a lower mass halo.
The HST Cycle 22 program for 6 QSO sightlines and 11 galaxy clusters at $z<0.4$ (PID 13833, PI Tejos) will provide an additional data-set for the study of \lya absorbers in clusters.

\subsection{Cluster \lya Absorbers and Galaxies}

In some cases the \lya absorbers can be associated with particular galaxies that are nearby in position-velocity space, rather than simply the large scale IGM. For Coma, the three absorbers within \rvir\ and $1\sigma_{\rm v,Coma}$ along the Ton0694 sightline (see Figure~\ref{v_gal.fig}) appear to be in the halos of galaxies given they are within 300 kpc and 300 \kms~of known galaxies.  Given their location within \rvir\ and $1\sigma_{\rm v,Coma}$, it is likely these three absorption lines represent gas that has been stripped from the galaxies by the hot ICM \citep[e.g.,][]{Gunn1972a,tonnesen10}.  The galaxies are relatively faint and belong to the red sequence, so they appear to already be quenched.  
Of the 3 absorbers, the one with the most nearby galaxies at 6210 \kms~is the highest column density absorber in our sample and has associated metal absorption lines.  The majority of the metal lines detected in the Virgo study also have galaxies within 300 kpc and 300 \kms with the NED sample, and these are likely to originate from galaxies.   However, for Virgo we found that those galaxies within \rvir~are largely stripped of halo gas \citep{Yoon2013a}, so the Coma Ton0694 sightline is unique in tracing galaxies that have halo gas and appear to be within the cluster based on their position/velocity location. This result may be linked to the infalling sub-structures in Coma's central regions \citep{Colless1996a,Neumann2001a,Brown2011a}.  Coma has interesting new results on infalling small galaxies that may also be relevant \citep{vandokkum15}.

Besides the three absorbers along the Ton0694 sightline, the Coma absorbers are not found to have galaxies within 300 kpc and 500 \kms~with the NED galaxy sample. This may change if fainter galaxies are included in the comparison.  For Virgo, at 17 Mpc, we find a large number of the absorption lines beyond \rvir\ and $1\sigma_{\rm v}$ are near galaxies.  Coma's lack of galaxy-absorber proximity may hold however, as the Virgo sightlines tend to fall in regions of higher galaxy density (Figures~\ref{coma_plot.fig} \& \ref{virgo_plot.fig}), and these denser galaxy regions do not change with magnitude restrictions.   Coma shows elongation in the north-west direction, but we do not have sightlines along this axis.

\subsection{Absorbers as Tracers of Turbulent Gas Motions}

Cluster masses are key for the derivation of several cosmological parameters and are obtained using x-ray observations and assuming hydrostatic equilibrium (see \cite{nagai14} for an overview).
In the outskirts of galaxy clusters, the assumption of hydrostatic equilibrium may break down due to turbulence from infalling groups and IGM streams \citep{zinger16,nelson14,nagai07}.  The kinetic energy from the turbulent motions provides additional support against gravity and can lead to 10-30\% uncertainty in the cluster mass estimates.
The clumps of warm gas detected by our observations are likely representative of some of the turbulent phenomena that can cause deviations between observations and hydrodynamical simulations \citep{Nagai2011,lau09,evrard96}.  
For the Virgo Cluster, the decrease in temperature in X-ray observations from 450 kpc - 1.2 Mpc is consistent with the presence of cool clumps of gas at large radii \citep{urban11}.  For the Coma Cluster, the measurements are not yet conclusive on the shape of the profile in the outer regions \citep{simionescu13}, though turbulence is expected \citep{zuhone16}.
Based on the position-velocity locations of our \lya absorbers, we expect the vast majority of the cool gas we detect to lie beyond the virial radii of the clusters;  still, the dispersion of the galaxies in the entire region compared to that of the absorbers provides a constraint on the level of turbulent motion in the outskirts.

 For Coma, the population of \lya absorbers has a higher dispersion than that of the abundant galaxies (see Figure~\ref{vdist.fig} and \S3.1).  This is indicative of additional motions within the gas and turbulence is likely to contribute.  The dispersion for the galaxies in and around Coma in the region of Figure~\ref{coma_plot.fig} is 1031$+/-19$\kms, which is similar to previous estimates of Coma's velocity dispersion \citep[1008\kms][]{Struble1999a}, and is consistent with the cluster being relatively dynamically relaxed.  In contrast, the absorbers have a dispersion of 1626$+/-236$\kms, and this is suggestive of deviations from hydrostatic equilibrium.  Unfortunately, for the closer Virgo Cluster, approximately half of the velocity range of the cluster is obscured and the dispersion cannot be reliably determined.   
    It is expected that the gas motions would be even higher in more unrelaxed systems, and this should be investigated with other clusters for which the entire velocity range can be sampled.  The results presented here suggest the gas traced by \lya~absorbers has an additional turbulent component compared to the dark matter dominated galaxies.
 
\subsection{Comparison between Cluster Simulations and Observations}

The recent simulations of EBP15 have examined the properties of warm gas probed by \lya absorbers in and around a Virgo-like and Coma-like cluster. In this section we use these simulations as a comparison point to our observations to gain insight into the origin and physical properties of the \lya absorbers.  As expected for dark matter halos of this mass scale, in the EBP15 simulations the vast majority of the gas flowing into the cluster is shock heated to temperatures $>10^6$~K.   The gas identifiable as \lya absorbers generally has a temperature below $10^5$~K.
It represents only a few percent of the total amount of gas within the virial radius, but approximately 1/3 of the gas beyond R$_{\rm vir}$.  This is consistent with the absorbers largely representing the IGM flowing into the cluster.  The material stripped from galaxies as they fall into the cluster are traced by higher column density absorbers with associated metals and these are rarer and often found within the virial radius.

With the infalling IGM and galaxies, the expected distribution of the absorbers in the simulations is to have most at low column densities ($<10^{15}$ cm$^{-2}$) beyond the virial radius, while the higher column density absorbers are more commonly found within the virial radius. The relevant quantity to compare to the observations is the median column density of the absorbers in projection. For the Virgo-like simulated cluster, this results in the median column density close to $10^{13}$ cm$^{-2}$ beyond \rvir, but approaching $10^{15}$ cm$^{-2}$ within 0.5 \rvir.  The more massive Coma-like simulated cluster does not show this increase due to the more abundant IGM flows into the massive cluster and most of the galaxies already being stripped. For the Virgo observations, we find primarily low column density absorbers throughout, and if anything the higher column density absorbers beyond \rvir.  The difference for the observed and simulated Virgo Cluster, may be due to the irregular structures of the clusters not matching exactly and a spherical virial radius analysis for an irregular cluster not being appropriate.  For the Coma observations, we find a relatively flat distribution of column densities as expected and that the highest column density absorbers are within \rvir~and consistent with being stripped galaxy material (see Figure~\ref{Rclt_W.fig}).  The latter may represent that the simulated Coma cluster is more settled than the actual Coma Cluster.
 As with most comparisons, additional observed sightlines would be highly beneficial.

Since the warm gas flows into a galaxy cluster from several directions in the simulations and is shock-heated at approximately the virial radius, the absorbers not only avoid the central part of the cluster in space, but also in kinematics.  The effect is the simulation predicts that most absorbers would lie within the velocity dispersion of the cluster, but have a bimodal velocity distribution.  This is not seen for the Coma Cluster (see Figure~\ref{vdist.fig}). This may be partially due to the limited number of sightlines, but thus far there is an even distribution across the velocity dispersion of the cluster and beyond. The fact that the galaxies clearly peak at the central velocity of the cluster while the absorbers do not may be consistent with a deficit of absorbers at the central cluster velocities. The Virgo absorbers may show a deficit at the central velocities of the cluster, but we are only able to detect absorbers on one side of the cluster's systemic velocity.     

The expectations from the simulations are that the more metal-rich, higher column density absorbers represent material stripped from galaxies. 
We have detected metal lines associated with three Ly$\alpha$ absorbers in the Coma Cluster and six \lya absorbers in the updated catalog for the Virgo Cluster. 
The metal lines are detected primarily for high column density absorbers, i.e., 2 out of 3 metal lines for Coma and 5 out of 6 for Virgo are for \lya~absorbers with \nhi~$>10^{14}$~\cm-2, though this could largely be a selection effect.  For Coma, two of the three \lya absorbers with associated metal lines are within 500 kpc and 100 \kms\ of a galaxy.  For Virgo, all but one of the \lya absorbers with metal lines has a galaxy within 300 kpc and 300 \kms.
Therefore there is overall consistency between the simulations and observations in the context of metal lines and there association with galaxies.

EBP15 found a lower covering fraction in the simulations compared to the Virgo Cluster results presented in Y12.  This discrepancy remains with the addition of the Coma Cluster and the new Virgo Cluster sightlines both within and outside \rvir.  If the observations are cut to only include absorbers within $1\sigma_{\rm v}$, the Coma-like simulation has a lower covering fraction than the Coma observations by a factor of 3-10 (depending on the column densities considered and relation to \rvir), and the Virgo-like simulation has a covering fraction discrepancy primarily as being lower at $>$\rvir (though this is for only half of the velocity range).   Since the simulations do have absorbers extending out beyond $1\sigma_{\rm v}$, the $2\sigma_{\rm v}$ cut may be more appropriate and in that case the Coma observations have a covering fraction $\sim6-20$ times higher and the Virgo observations have a similar discrepancy.  Given the limited number of sightlines in the observations and the difficulty in accurately simulating column densities, additional simulations and observations will help to clarify this discrepancy.

In summary, though there are some discrepancies between the simulations and observations, there is overall agreement with the observations having a combination of inflowing IGM and galaxies and showing a distinct relationship between impact parameter and column density for the highest mass halos.

\section{Summary}

This paper presents the distribution of warm gas surrounding two massive dark matter halos, the Coma and Virgo Clusters.
We have detected 14 \lya absorbers within $3\sigma_{\rm v}$ and 2\rvir~of the Coma Cluster using COS on HST.  We also present an update to Y12 for the less massive Virgo Cluster resulting in a total of 
49 \lya absorbers surrounding this cluster (see \S3 for details).   The absorbers span a range of column densities, but the majority are at $<10^{14}~\cm-2$.

The distribution and column densities of the absorbers indicate the majority representing the infalling IGM, with a smaller number being stripped galaxy material.  The lack of absorbers within 1\rvir~and $1\sigma_{\rm v}$ is consistent with the gas being shock-heated within the virial radii of clusters.
Specific results to highlight can be summarized as follows.

\begin{itemize}
    \item In contrast to the smaller dark matter halos of galaxies, the relationship of increasing column density with decreasing impact parameter is not observed for the clusters.  In addition, the covering fraction is lower within \rvir~and somewhat higher beyond \rvir~for the galaxy clusters than for individual galaxies.
    
    \item The velocity dispersion of the absorbers is higher than that of the galaxies for the Coma Cluster.   This is indicative of additional motions acting on the gas and is consistent with the absorbers tracing turbulence in the outskirts.
    
    \item Along one sightline in the central part of the Coma Cluster the absorbers can be associated with nearby galaxies.  For the Virgo Cluster, the newly detected metal lines are in close proximity of galaxies.  Future studies of the relationship between galaxies and absorbers in different environments using a volume limited sample would aid our understanding of galaxy quenching.
    
    \item There is overall agreement between the observations and the EBP15 simulations of a Coma-like and Virgo-like cluster.  There is a combination of infalling IGM and stripped galaxy material and the spatial and kinematic distributions of the observations and simulations are consistent, though additional observational sightlines are needed.   The primary exception is the covering fraction of the warm gas is higher in the observations than the simulations.
    
\end{itemize}

\begin{acknowledgements}
We thank Greg Bryan, Joe Burchett, Nicolas Tejos, John Stocke, Ben Oppenheimer and Xavier Prochaska for useful discussions, Charles Danforth for sharing his COS coaddition code and Paola Serra for sharing his HI data for NGC 4921.  This project was funded through support for HST programs 13382 and 13383
from NASA through a grant from the Space Telescope Science
Institute, which is operated by the Association of Universities for
Research in Astronomy, Inc., under NASA contract NAS5-26555.  MEP acknowledges support from Dan and Elaine Putman and the University of Colorado for hosting me during much of this work.
\end{acknowledgements}

\bibliographystyle{apj}
\bibliography{ms}

\end{CJK*}
\end{document}